\documentstyle[12pt]{article}
\title{Production and decay of single heavy spin-3/2 leptons in high energy
electron-positron collisions}

\author{F. M. L. Almeida Jr., J. H. Lopes,\\
 J. A. Martins Sim\~oes and A. J. Ramalho\\
Instituto de F\'{\i}sica\\
Universidade Federal do Rio de Janeiro\\
Ilha do Fund\~ao, Rio de Janeiro\\
BR-21945-970, RJ, Brazil}
\date{}
\begin{document}
\maketitle
\begin{abstract}
    We discuss the possibility of detecting heavy spin-3/2 leptons at LEP2
and NLC energies. The angular distributions of
primary ordinary leptons are shown to be a good measure to discriminate
between a heavy spin-3/2 lepton and a similar nonstandard spin-1/2
signal.
\end{abstract}
\newpage
\par
    There is now an extensive amount of data that confirm the standard
model as the theory of fundamental interactions. However, since it depends
on a relatively large number of parameters adjusted by hand, there is an
expectation that some extended theory will provide a deeper understanding
of nature.  As yet none of the theoretical proposals has achieved this
goal. On the experimental side, a lot of effort has been put on the search
for new particles, which could shed some light on the physics beyond the
standard model. In this paper we study the production and decay of possible
heavy spin-3/2 leptons at LEP2 energy $\sqrt{s} = 190GeV$ and at a future
new linear collider (NLC) at $\sqrt{s} = 500 GeV $.

\par
The motivation for spin-3/2 leptons comes from at least three scenarios.
In composite models ordinary spin-1/2 leptons and quarks are thought to be
formed by more fundamental fermions. If this hypothesis is correct, then
excited spin-3/2 states are expected to exist \cite{Lopes80}. Composite
fermions in enlarged groups ( relative to the standard  group) also imply
spin-3/2 fermions \cite{Tosa85}.  In supergravity gauge theories
\cite{Nieu81} the spin-3/2 gravitino is the fermionic counterpart to the
gravitational field. More recently, spin-3/2 fields  appear  in a
quaternionic quantum mechanical model\cite{Adler94} for composite leptons
and quarks. Unfortunately, there exists no complete self-consistent theory
of interacting spin-3/2 fields. At the classical level there is the
problem of faster than light propagating modes \cite{Velo69}. The minimal
electromagnetic coupling for pair production of spin-3/2 leptons violates
unitarity just above threshold \cite{Tayl80}. These problems make
difficult a comparison of theoretical predictions for spin-3/2 objects
with any possible "new physics" experimental signal. Some time ago it was
pointed out \cite{Lopes80} that some of these difficulties could be
ameliorated for the coupling of spin-3/2 fermions to massive electroweak
gauge bosons. Although this hypothesis is not a solution of the problems
concerning spin-3/2 fields, one expects that it will allow that reliable
phenomenological estimates be made, in the same spirit of the effective
Fermi theory for weak interactions.  The heavy gauge boson propagator can
act as an effective form factor, and one can compute production cross
sections for spin-3/2 fermions which  violate unitarity in kinematical
regions well above thresholds, as in the old Fermi model.  There are some
theoretical arguments for a high spin-3/2 mass. The well known
Weinberg-Witten theorems \cite{Wein80} leave open the possibility of
massive spin-3/2 states at higher energy scales.  This is
supported, to some extent, by a recent analysis of the LEP1 data carried
out by Montero and Pleitez \cite{Mont94}, which implies the bound
$M_{3/2} > M_Z$.  We are then lead to study
massive spin-3/2 fermions with masses above the Z mass.
\par
The phenomenology of spin-3/2 lepton production has been studied for
electron-positron collisions \cite{Lopes80}, hadronic reactions
\cite{Speh87} and more recently, for the electron-proton collider HERA
\cite{Alm93} and the $e^-\gamma$ mode of the next generation of
electron-positron colliders
\cite{Eboli95}.  In this paper we turn our attention to the production and
decay of a single spin-3/2 lepton in $e^+e^-$ collisions at LEP2 and NLC
energies. Spin-3/2 lepton pair production is also possible, but it is
bounded by $M_{3/2} < \sqrt{s}/2$, whereas single spin-3/2 lepton production
can reach masses as high as $\sqrt{s}$.
  We compare our results for spin-3/2 with the analogous heavy
spin-1/2 case.  A detailed analysis of the production process of single
heavy spin-1/2 leptons was carried out in reference \cite{Alm95}, in the
framework of different nonstandard electroweak models, and the model
dependence of unpolarized production cross sections and distributions
 was found to be weak.

\par
          In order to reduce the model dependence on spin-3/2 couplings,
we consider the following phenomenological currents:
\begin{equation}
J_1^{\mu} = e {\bar u}^{\mu} (p,3/2)
 (c_{1V} - c_{1A} {\gamma}^5) u(k,1/2)
\end{equation}

\begin{equation}
J_2^{\mu} = {e\over \Lambda}{\bar u}^{\lambda} (p,3/2) q_{\lambda}
{\gamma}^{\mu} (c_{2V} - c_{2A} {\gamma}^5) u(k,1/2),
\end{equation}

\noindent
where ${\bar u}^{\mu} (p,3/2)$ represents a Rarita-Schwinger vector
spinor, $q^{\mu} = p^{\mu}-k^{\mu}$ and $\Lambda$ is a mass scale that
sets the strength of the interaction. We assume that the mixing between
ordinary spin-1/2 leptons and heavy spin-3/2 leptons occurs only within
the same lepton family, with conservation of lepton number. A more general
classification of operators with dimension up to six,
involving spin-3/2 fields is given by Burges and Schnitzer \cite{Burge83}.

\par
    A spin-3/2 heavy lepton whose mass exceeds the experimental lower
bound $M_{3/2} > M_Z$ generally decays via a two-body process. The widths
of the spin-3/2 leptons, according to currents $J_1$ and $J_2$, are given
by
\begin{equation}
{\Gamma}^{(1)} = {\alpha \over 48}(c_{1V}^2+c_{1A}^2) M_{3/2}
{(1-\kappa)^2 \over \kappa} (1+ 10\kappa + {\kappa}^2)
\end{equation}
\noindent
and

\begin{equation}
{\Gamma}^{(2)} = {\alpha \over 48}(c_{2V}^2+c_{2A}^2) M_{3/2}
({M_{3/2} \over \Lambda})^2{(1-\kappa)^4 \over \kappa} (1+2\kappa) ,
\end{equation}

\noindent
where $\kappa = {(M_{3/2}/M_B)}^2$ and $M_B$ stands for the vector boson
(W or Z) mass. For $\Lambda = M_{3/2}$ and coupling constants of the order
of one, typical widths are then in the range $0.1-10MeV$, showing
that the direct observation of these leptons is a hopeless task.  The
branching ratios for two-body decays are given in table 1. For the charged
spin-3/2 lepton the most favored signature will be  one ordinary charged
lepton (primary lepton) in one hemisphere and a W boson and a
missing $\nu$ in the other. For the neutral case the dominant event
 will have one missing neutrino in one hemisphere, along with a W boson and
charged ordinary lepton in the opposite hemisphere.

\par
It is well known that there is no theoretical prediction on the absolute
value of the total cross section for spin-3/2 production. We have then
normalized the total cross section to the upper bound on similar spin-1/2
production \cite{Alm95}. In Fig.1 we show the Feynman diagrams for single
spin-3/2 lepton production. The explicit calculation of these diagrams
leads to
\noindent

\begin{eqnarray}
{{d\sigma}\over dx}^{(1)}&=&{\pi \over 12} {\alpha}^2
{(1-\eta) \over M_{3/2}^2}^2
 \Big\{  f^{(s)} A_1 \left \lbrack 8\eta + (1-\eta)^2 (1-x^2)
-4B_1 \eta x \right \rbrack \nonumber\\
&&+ f^{(t)} \lbrack A_1 (1+\eta) (1+2\eta)
+ B_1 \eta (3-\eta) - 2 (A_1 + B_1) \eta x\nonumber\\
&& \hskip 1cm - (1-\eta) (A_1 (1-2\eta) + B_1 \eta ) x^2 \rbrack \nonumber\\
&&+ f^{(st)} (A_1+B_1) \left \lbrack 4\eta +
(1-x) (4\eta + (1-\eta)^2 (1+x)) \right \rbrack \Big\}\\
\nonumber\\
{{d\sigma}\over dx}^{(2)}&=&{\pi \over 96} {\alpha}^2
{(1-\eta)^2 \over {\eta {\Lambda}^2}}
\Big\{ 8 f^{(s)} (1-\eta)^2 \left \lbrack A_2 (1+\eta+(1-\eta)x^2)
+2B_2 x \right \rbrack \nonumber\\
&&+ f^{(t)} {\left \lbrack 1+\eta-(1-\eta)x \right \rbrack}^2\nonumber\\
&&\hskip 1cm \times \left \lbrack (5+\eta)A_2 -
(3-\eta)B_2 +(A_2+B_2)(2x+(1-\eta)x^2) \right \rbrack \nonumber\\
&&-4 f^{(st)}(1-\eta) (A_2+B_2)(1+x) \nonumber\\
&&\hskip 1cm \times \left \lbrack 1+\eta +(1-\eta)x
\right \rbrack \left \lbrack 1-\eta -(1+\eta)x
\right \rbrack \Big\}
\end{eqnarray}
\\

In the expressions above, $x = \cos{\theta}$, where $\theta$ is the angle
of the primary lepton with respect to the incoming electron, and the coupling
constants $g_{iV}$ and $g_{iA}$ refer to standard-model vertices of the
form
$\displaystyle{-ie{\gamma}_{\mu}(g_{iV}-g_{iA}{\gamma}_5)}$. In addition,
$$\displaystyle{\eta = M_{3/2}^2/s},$$
$$\displaystyle{f^{(s)} = {s^2 \over {(s-M_Z^2)^2+({\Gamma}_Z M_Z)^2}}},$$
$$\displaystyle{f^{(t)} = {s^2 \over (t-M_Z^2)^2}},$$
$$\displaystyle{f^{(st)} = f^{(s)}{(s-M_Z^2) \over (t-M_Z^2)}},$$
$$\displaystyle{A_i = (c_{iV}^2+c_{iA}^2)(g_{iV}^2+g_{iA}^2)}$$ and
$$\displaystyle{B_i = 4c_{iV}c_{iA}g_{iV}g_{iA}},(i=1,2).$$
\par
    Fig. 2 shows the production cross sections for charged spin-3/2
leptons, both at $\sqrt{s} = 190GeV$ and $\sqrt{s} = 500GeV$. For the sake
of comparison, the cross section for the production of a heavy spin-1/2
lepton is also shown for a model with mixing angles equal to ${\theta}_i =
0.1$. Assuming that LEP2 and NLC luminosities will take
on the conservative yearly values of $150 pb^{-1}$ and $10 fb^{-1}$
respectively, the production rate of a $120GeV$ spin-3/2 lepton is found
to be around 4 events per year at LEP2 but may reach 700 events per year
at NLC. The energy dependence of the total cross sections is displayed in
Fig.3. The curves grow steeply with the center of mass energy, which
suggests the breakdown of unitarity. As expected, this is more severe in
the case of the higher-dimension operator $J_2$. However, this violation
of unitarity occurs for energies far above the production threshold.

\par
    Clearly, it would be difficult to separate an eventual spin-3/2 heavy
lepton from a corresponding spin-1/2 signal only on the basis of total
cross sections. Therefore it is natural to investigate which
distributions are useful to determine the spin of nonstandard heavy
leptons, and/or to distinguish the phenomenological currents $J_1$ and
$J_2$. The results of this study are summed up in Figs.4-6, which show
the normalized angular distributions of the charged ordinary lepton that
accompanies  the charged heavy lepton. In Fig. 4 the structure of the
phenomenological  currents $J_1$ and $J_2$ was assumed to be $V \pm A$,
V or A. One can see that there is no significant dependence on the coupling
constants $c_{iV}$ and $c_{iA}$. In figs. 5-6 we compare the spin-1/2 and
spin-3/2 cases. The primary charged lepton associated with the heavy spin-1/2
charged lepton is produced mostly in the forward hemisphere, especially at
NLC energies. As a matter of fact, this behavior is reproduced in the case
of spin-3/2 current $J_1$ when the lepton mass is large enough, but it may
still be possible to distinguish the two angular distributions if
polarized $e^+e^-$ beams are employed, along the same lines as suggested
in \cite{Alm95}. If heavy spin-3/2 lepton production is correctly described by
current $J_2$, however, the probabilities of detecting the primary lepton
in each hemisphere are roughly the same at LEP2 energies. Although this is
no longer true at NLC energies, the normalized angular distribution which
corresponds to current $J_2$ remains different from the other two.
\par
    In conclusion, we have shown that it is feasible to search for
new spin-3/2 heavy leptons at LEP2 and NLC colliders, and that in most cases
their experimental signals can be discriminated from similar spin-1/2 effects.

\par

{\it Acknowledgments:} This work was partially supported by CNPq, CAPES,
FUJB, FAPERJ and FINEP.

\newpage
\begin{tabular}[t]{llcr}
\hline
\\
$L^-_{3/2}\longrightarrow (l^-\nu_l)\ \nu$ &
$L^0_{3/2}\longrightarrow (l^+\nu_l)\ e^-$ & & \ \ \ \ 8
\\
\\
\hline
\\
$L^-_{3/2}\longrightarrow (hadrons)\ \nu$ &
$L^0_{3/2}\longrightarrow (hadrons)\  l^-$ & & \ \ \ \ 50
\\
\\
\hline
\\
$L^-_{3/2}\longrightarrow (invisible)\ e^-$ &
$L^0_{3/2}\longrightarrow invisible$ & & \ \ \ \ 5
\\
\\
\hline
\\
$L^-_{3/2}\longrightarrow (l^+l^-)\ e^-$ &
$L^0_{3/2}\longrightarrow (l^+l^-)\ \nu_L$ & & \ \ \ \ 3
\\
\\
\hline
\\
$L^-_{3/2}\longrightarrow (hadrons)\  e^-$ &
$L^0_{3/2}\longrightarrow (hadrons)\  \nu_L$ & & \ \ \ \ 18
\\
\\
\hline
\\
\end{tabular}

\noindent Table 1:Branching ratios (\%) for the two-body
 decays of heavy\\ spin-3/2 leptons. The parentheses show the
 products of the $W$\\ (or $Z$) boson decay.

\newpage

\section*{Figure captions}
\newcounter{ijk}
\begin{list}%
{Fig.\arabic{ijk}\ }{\usecounter{ijk}\setlength{\rightmargin}{\leftmargin}}

\item The lowest order Feynman diagrams for the process $e^+e^-
\longrightarrow  e^-({\nu}_e) L^+({\bar{L}}^0)$.

\item (a) Cross sections for single production of heavy spin-3/2
leptons at $\sqrt{s} = 190GeV$. The cross section for a nonstandard spin-1/2
lepton of a vector doublet model is shown for comparison. (b) Same as (a),
but at $\sqrt{s} = 500GeV$.

\item Total cross sections as functions of the center of mass
energy, for input masses $M_L = 120GeV$ and $M_L = 400GeV$.

\item Normalized angular distributions of the primary lepton for different
couplings $(V \pm A, V or A)$ in spin-3/2 currents $J_1$ and $J_2$.

\item (a) Normalized angular distribution of the primary lepton
in heavy spin-3/2 lepton single production, at LEP2 energies and for an input
mass $M_L = 120GeV$. The corresponding distribution for the case of
heavy spin-1/2 lepton production is also shown. (b) Same as (a), but for
an input mass $M_L = 160GeV$.

\item (a) Normalized angular distribution of the primary lepton
in heavy spin-3/2 lepton single production, at NLC energies and for an input
mass $M_L = 250GeV$. The corresponding distribution for the case of
heavy spin-1/2 lepton production is also shown. (b) Same as (a), but for
an input mass $M_L = 400GeV$.
\end{list}

\end{document}